# Developer Load Normalization Using Iterative Kuhn-Munkres Algorithm: An Optimization Triaging Approach

Madonna Mayez
Software engineering
*British University in Egypt*
Egypt
Madonna.mayez@bue.edu.eg

Khaled Nagaty
Computer Science
*British University in Egypt*
Egypt
Kaled.nagaty@bue.edu.eg

Abeer Hamdy
Software engineering
*British University in Egypt*
Egypt
Abeer.hamdy@bue.edu.eg

*Abstract*— Bug triage can be defined as the process of assigning a developer to a bug report. The duty of the bug triager is to study the developers' profiles well in order to make an appropriate match between the developers and the incoming bug reports. Thus, this process is a vital step in issue management system. In fact, the number of bug reports submitted every day is gradually increasing which affects the developer workload. Thus, the triager should consider this factor in distributing the bugs and because of the manual approach, many developers are burden. In particular, triaging bug reports without considering the workload does not only affect the developers' workload but also leads to an increase in the number of unaddressed bug reports. As a result, the fixing time of the reported bugs will relatively increase. Unlike other researchers who focus on automating the bug triage and ignoring the developer workload, in this work, we handle the triaging process from a different perspective. The proposed approach focuses on how to optimize the bug fixing time by normalizing the developer load in an automating system. To evaluate our work, we use 26,317 bug reports from different bug repositories. Results shows that our work outperforms other systems in terms of optimizing the bug total fixing time and normalizing developer load.

## I. Introduction

The usual scenario is the end user come upon a bug while testing or working on a software and document the issue in a bug tracking system using the bug report document [1]. Issue tracking systems (ITS) such as Bugzilla [2] and Jira [3] are interdisciplinary domains, which used to organize and track the change requests and bug reports. One of the main steps of handling the submitted bugs is the triaging process. Triaging bugs means assigning each bug to the most appropriate developer [4]. So, the existing scenario is the triager manually studies the developer profile and try to match between the skills needed in the bug reports and what a developer has according to his history.

A bug report takes a long journey starting from submitting on the system till being solved. In each stage a bug report has a different status [5]. Once a bug report is submitted, its status is set to NEW. When the triager assigns it to a developer, its status will be changed to ASSIGNED. If the fixer successfully fixed it, the status will be changed to CLOSED. Figure 2 describes the bug report life cycle. The bug status is set to DUBLICATE, when a bug is marked as duplicated. When a bug will not be fixed, it is marked as not actual bug and its status is set to INVALID [6]. The VEREFIED bugs are solved by the developers and reviewed by the mangers, then they will be ready to be CLOSED. They will not be opened again.

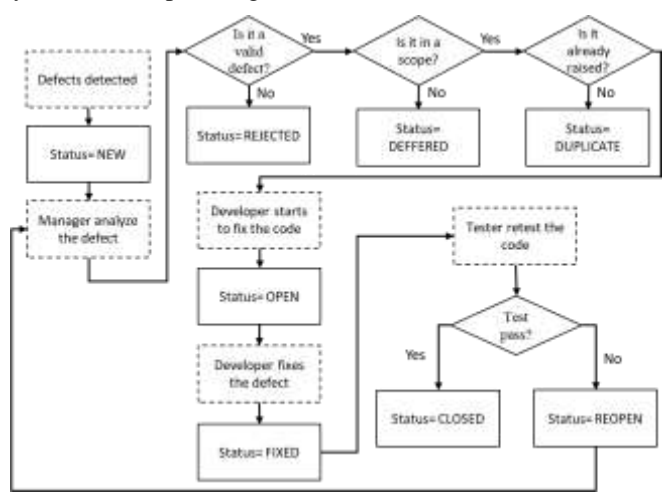

Figure 1: Bug report life cycle [5].

In large-scale systems, the number of bug reports is increasing day by day. For example, in November 2019, Eclipse ITS recorded 500,000 issue report and Mozilla ITS received more than 1,600,000 ones [7]. Thus, this process could be error prone. In fact, around 37% of the submitted reports are tossed for fixing [8]. Tossing a bug report means reassigning it to another developer because the original one failed to fix it under some conditions. These tossing actions should be limited as much as possible. It not only waste financial and human resources, it also delays the bug fixing time. For example, if a sever bug report is assigned to a developer with less qualifications, there are two possible scenarios. First, the developer can be loaded with other reports and will take much time to address the problem. Second, the developer will not be able to handle it and it will be reassigned to another one. Both will require longer time to handle the problem and accordingly, more human and financial resources will be needed.

Most of the current Studies are conducted to explore such decision making using different approaches such as machine learning, text mining and information retrieving techniques. However, most of them take the automation direction using traditional classifiers. They consider the developers as classes and the bug reports are the entities. This definitely reduce the time but it keeps the fixer load ignored. Besides, researchers claim that the manual assignments are not accurate enough, however, they use it as a baseline to evaluate their results. Therefore, we go slightly deeper and try to fix it as an optimization problem.

The paper aims at answering the following research questions:

Q1: How triaging problem can be represented as a bipartite graph?

First, we model the triaging problem as a task-resource allocation problem because there should be a developer for each bug report. So, we use the bipartite graph matching [9] [10] to formulate the problem. The two sets on both sides are the bug reports and developers. Thus, the problem can be modeled as a complete bipartite graph, because each node in S1= {BR1, BR2, BR3…} has a relation (edge) to be linked with each node in S2= {D1, D2, D3…}. This link shows the average time that each developer may spend to fix the corresponding bug report.

Q2: what is the effect of topic modeling in developer scoring system?

The role of topic modeling is to differentiate between the bug reports (labeling stage). Therefore, each bug report is classified by specific topic. Accordingly, the developer skills will be represented as a score per developer. For example, if the bug reports are classified into 5 topics, the developer score will be a vector of length 5, a score in each topic.

Q3: How does the iterative- Kuhn-Munkres approach optimize the total fixing time and the developer load?

Iterative Kuhn-Munkres is an optimization system that can select a group of developers who can fix a group of bugs in shorter time taking into consideration the developer workload. In each iteration a group of bug reports will be distributed on the developers equally according to the type of bug report and the score of developer. The number of Iterations equals to the number of testing data divided by the number of developers.

Q4: what are the evaluation matrices the iterative- Kuhn-Munkres approach is evaluated by?

Based on an experimental analysis, the proposed technique is evaluated by how many hours the recommended developers may take to handle set of bugs against the number of hours that the real developers took to fix the same set. In addition, the developer workload in the proposed approach against in reality.

## A. Motivation

The motivation behind using the Kuhn-Munkres algorithm is the great optimization effect such as what recorded by X.Zheng in [11], he optimized the search efficiency (23.4%–78.1%) and many other researchers who proved the effectiveness of the Kuhn-Munkres algorithm in cost optimization for assignment problems [12] [13]. In addition, Yang [14] inspired us by the use of LDA to get the important topics out of the historical bug reports and we extended this idea by using it as labels for each bug. Although, many researchers used various technique to speed up the triaging process, very few of them considered the human factor (developer workload). In this paper, we present a framework to fill this gap using Munkres algorithm and LDA.

Paper Organization: The rest of this paper is organized as follows. Section II documents the related works with the related techniques and the research gap in the existing bug triage approaches. Section III introduces the proposed algorithm and including the fundamental concepts and experiment setup. Results are stated and discussed in section IV. Finally, V is a conclusion.

## II. LITERATURE SURVEY

To optimize the bug fixing process, researchers focused on automating the triaging task. They used different approaches, namely information retrieval (IR) based [4] [15] [16] [17] [18], machine leaning (ML) based [19] [20] [21] [22] [23] [24] [27], and graph theory-based approaches. We will give a brief introduction on each one of them.

### A. Information retrieval techniques

Matter [4] used vector space model to extract information from bug reports source code contributions for the purpose of modeling a developer's expertise. In the triaging process, authors compared the vector of the upcoming new bug vector with the vectors of developers' expertise. They used bug reports of eight years from Eclipse [25] development bug data. They included 130,769 bug reports for the case study. Using top-10 recommendations list, they achieved triaging accuracy up to 71.0%

Sahu [15] proposed a hybrid selection method to reduce the database size, feature selection, and Instance selection methods to reduce the number of features used from the bug reports dataset. He proposed a data reduction technique based on a combination of KNN+NB approach. By evaluating this method on Mozilla dataset [26], results show accuracy percentage 85%.

Xin [16] proposed a new incremental approach for automating the bug triage process named TopicMinerMTM which extends Latent Dirichlet Allocation (LDA) algorithm [27]. It computes to what extent each developer is matched to the new bug report. It takes into account the topic distribution of a new bug report to be assigned to the most appropriate fixer based on the affinity of the fixer to the bug report topics considering the topics in the fixer's history. They evaluated the TopicMinerMTM approach on GCC [28], OpenOffice [29], Mozilla [26], NetBeans [30], and Eclipse [25]. The total number of bug reports used is 227,278 bug reports. The results show the prediction accuracy is between 48% to 90%.

Sun [17] used the approach of removing the duplicated bugs for the purpose of optimizing the time and effort in the bug triaging process. So, instead of two fixers are busy by the same bug their approach is an automatic detector for the duplicated bugs. They build their technique based on a retrieval function REP, which extends the BM25F (similarity formula used in [31])to take into account the features in long bug reports such as reported product, component, and version. They used Mozilla [26], Eclipse [25] and OpenOffice [29] to validate their technique. The results show 10–27% relative improvement in recall and 17–23% relative improvement in mean average precision. They applied the technique on large dataset, 209,058 reports from Eclipse. The result was 37–71% in recall and 47% in mean average precision.

Nguyen [18] also take the approach of Sun for reducing the cost of bug triaging by detecting the duplicated bugs. He extended Sun's technique by combining information retrieval (IR)-based features and topic-based features in an approach called duplicate bug report detection approach (DBTM) using the same dataset used in Sun's work. It models a bug report as a textual document describing one or more technical issues/topics in a system to address textual dissimilarity between duplicate reports. For the purpose of feature extraction, they combined both IR and topic modeling by applying ensemble averaging technique, which is defined as the mean of quantity. They used the Gibbs sampling to train DBTM on the dataset with identified duplicate bug reports and later to identify duplicate ones. Gibbs sampling is a sampling technique which called Markov chain Monte Carlo (MCMC) algorithm that uses multivariate probability distribution to achieve sequence of observations. The evaluation shows that DBTM improves the state-of-the-art approaches by up to 20% in accuracy.

*B. Machine learning techniques*

ML techniques consider the bug triage problem as a classification problem. It mainly depends on the previously resolved bug reports to train the classifier and then predict the qualified developer for the new reported bug.

KASHIWA [19] is the first researcher focused on the developer workload. He believes that by distributing the work equally among the developers, the bug fixing time will be reduced automatically. He formulated his work as a multiple Knapsack problem in purpose of finding the best combination between the bug reports and fixers taking into consideration the load for each fixer. To apply his methodology, he used SVM classifier and Laten Dirichlet allocation. He evaluated his work using bug reports from Mozilla Firefox, Eclipse Platform and GNU compiler collection (GCC) [32]. In fact, he was able to reduce the bug fixing tie by 35%–41%, compared to the manual triage.

Weiß et al [20] used K-Nearest Neighbor (KNN) clustering algorithm to predict the bug fixing time. They proposed that predicting the bug fixing time depend on the fixing time spent on similar historical bug reports. They used text similarity for measuring the similarity between a new bug report and the historical ones. They concluded that their prediction model beats the naive approach.

Lucas et al [21] explored the capability of using data mining tools in predicting the bug report time fixing. They used 0-R, 1-R, Decision Trees, Naive Bayes and Logistic Regression. Using historical bug reports from Bugzilla [2] and Eclipse [25], they recorded Logistic Regression is the best which achieved 34.9% prediction accuracy.

S. Mani [22] proposed a novel bug report representation algorithm using deep learning model called deep bidirectional recurrent neural network (DBRNN-A) model. This model learns syntactic and semantic features from long word sequence. Among the bug report fields, he used the description and title (summary). He did a comparison among four different classifiers, namely multinomial naive Bayes, cosine distance, support vector machines, and softmax. He collected bug reports from three different repositories namely Google Chromium (383,104) [33], Mozilla Core (314,388) [26], and Mozilla Firefox (162,307) [26]. Through experimental results, he found that DBRNN-A along with softmax classifier is better than the bag-of-words model. Additionally, he concluded that the bug report description is important in improving the classification performance.

Bhattacharya [23] used title, description, keywords, products, components, last developer activity as information extracted from the bug report. They extracted the features using the bag of words technique in addition to the term frequency–inverse document frequency (TF-IDF), to simulate how a word is important in a corpus of document. For building the classifier for keywords extracted in reports, they used a ML approach with Naive Bayes, Bayesian network and tossing graph. Their dataset is quite large. They used 306,297 bug reports from Eclipse [25] and 549,962 from Mozilla [26]. They achieved up to 83.62% prediction accuracy in bug triaging. They reported 77.64% average prediction accuracy when considering bug report ID, which is the module a bug belongs to. However, the accuracy drops to 63% when bug report ID is not used.

Anvik [24] also used the ML approach for automating the triaging process. He did an innovative idea which is filtering the invalid and useless data based on some rules and constraints he put. While training the classifier, he ignored the no longer working or inactive developers and the unfixed bug reports. The information he used in the bug report is the title and description. For feature extraction he used the normalized tf which is a text mining technique to consider the term frequency for in a two-fold normalization. So, first each document is normalized to length-1. Then, By applying three different classifiers namely SVM, Naive Bayes, and C4.5, they achieved a precision of up to 64%.

Gondaliya [34] used natural language processing. He used different methods of text mining. First, lemmatization to reduce inflectional forms. Second, part-of-speech tagger (POS tagger) which is a software used to read texts in any language and accordingly, it assigns part of speech for each term. Third, bigram which consists of a sequence of n adjacent elements (n=2) from string to token for the purpose of providing the probability of a token given another token. The last one is stop-word removal to remove the stop words such as "a", "an". Then he applied classification algorithms to evaluate his work. He applied Linear Support Vector Machines (SVMs), multinomial naive Bayes, and Long Short Term Memory (LSTM) networks to classify the bug reports and accordingly triage them. He assessed his methodology using a small dataset (1215 bug report) and a large dataset (7346 bug report). In the small dataset, the accuracy for the three classification techniques was 47.9 → 57.2%. For the large dataset the performance ranged from 68.6% → 77.6%.

*C. Tossing graph*

TG approach is constructed based on the use of graph to identify the tossing (reassignment) probability between teams from the tossing history using the Markov chain models. After the triager assigns the bug report to the developer, he could be unable to solve it. Thus, the bug report will be reassigned (tossed) to another developer. This process will get into a loop till the bug reaches the bug resolver. Bug tossing increases the bug fixing time which wastes the bug triagers and developers time. According to Jeong [35], 37%-44% of bug reports in Mozilla [26] and Eclipse [25] are tossed to other developers. For instance, a bug report could be assigned by accident to unskilled developer, so it should be reassigned to the appropriate one.

Jeong [35] introduced a graph model based on Markov chains to handle the bug tossing process. This model provides developer network which can be used to discover the team structure and assigns the suitable developer to the new bug. They used 445,000 bug reports as the dataset from Mozilla [26] and Eclipse [25]. Experiments show that their model reduced the tossing rate by 72%. Moreover, it increased the prediction accuracy by up to 32%.

Yadav [36] proposed a novel strategy named developer expertise score (DES) which consists of three stages. The first stage is an offline process which assign a score for every developer based on priority, versatility and average fix-time for the contribution of each one. This process is about deciding the capable developers based on three similarity measures, namely feature-based, cosine-similarity and Jaccard. The second stage is about ranking the developers based on their DES. Third stage, they used three types of classifiers, namely Navies Bayes, Support Vector Machine and C4.5 paradigms to compare their work with the ML-based bug triaging approaches. Results show that, using five open-source databases (Mozilla [26], Eclipse [25], Netbeans [30], Firefox [26], and Freedesktop) with 41,622 bug reports, DES systems recorded a mean accuracy, precision, recall rate and F-score (to measure the tests accuracy) of 89.49%, 89.53%, 89.42% and 89.49%, respectively, reduced bug tossing length (BTL) of up to 88.55%.

*D. Fuzzy techniques*

Tamrawi [37] proposed an automated bug triaging approach called Bugzie, which depends on fuzzy set and caching the developers. His model cashes the developers in a queue according to their skills level and then the assignment action considers the first one in the queue first. It assigns a score for each developer based on his previously fixed bug reports. The purpose behind using the fuzzy technique, he assumed that a software system has many technical aspects, each of which is represented by a term in the fuzzy set. For a new bug report, his approach (Bugzie) combines the fuzzy set with respect to its terms and the developers score to predict the most appropriate developer. By evaluating this approach, Bugzie recorded 75-83% accuracy.

Feng Zhang [10] the impact of the time delays caused by developers on total bug fixing time. They performed their empirical study on three different opensource software systems (Mylyn, Eclipse Platform, and Eclipse PDE). They explored the factors that causes the delay along three dimensions (bug reports, source code and the required code changes to handle the bug). Their findings prove the real existence of the delay. They proved that their experiment can help development team in prioritizing bugs and optimizing bug assignment. So, they can reduce the delay to improve the bug fixing process. Additionally, they summarized the most influential bug report factors which are bug type, bug severity, Operating System, bug description bug comments, property of source code and property of code change.

Vinay et al [38] proposed the use of virtualization and lean methodologies [39] for optimizing the bug time fixing. They received good results through reducing the time from 15 hrs to 0.5 hrs saving 362.5 hrs for a team of 25 developers.

*E. Comparison to previous work*

In summary, most of previous work consider the triaging problem as a classification problem. So, they all took similar steps with some different algorithms in handling this problem. The accuracy percentage in each of them was not good enough. Besides, they claim that the manual triaging has a high probability to make mistakes which leads to many tossing actions. However, researchers also use the manual assignments as reference to evaluate their work. This is a contradictory step.
In this work, we are optimizing the triaging problem from a completely different perspective. We followed the mindset of KASHIWA [19] regarding how the normalized developer work load can make great effect on optimizing the bug total fixing time. We use the bipartite graph to represent the bug reports and developers as vertices and we will assign them using the Kuhn-Munkres algorithm which has a great advantage in handling the developer load.

*F. Research Problem*

In fact, triaging large number of bug reports manually by the triagger consumes time and resources. Large software development projects receive a daily huge number of reports and change requests. For instance, Mozilla project [26] received over 800,000 reports in October 2012, averaging 300 new change requests daily [40]. Each one of them require triaging. A triager, as a project member, must examine the recently submitted reports and start to make decisions about the process of assigning these reports among developers within the software development life cycle SDLC. The main key of triaging process is assigning the bug report to the developer who is able to make the necessary code change to fix the fault or implement the required new feature. Using these statistics, it is obvious that the triagger cannot distribute the bugs over the working developers equally. In addition, a bug report could have critical consequences on the project. Therefore, inaccurate decisions can increase the project cost significantly. Based on a survey done by the National Institute of Standards and Technology, the annual cost to handle software bugs is over $59.5 billion [41]. Besides, a senior bug fixer salary in Aka is $129,328 per year [42]. They are considered huge numbers.

### III. PROPOSED FRAMEWORK

The proposed system is built with intention to normalize the work load, the number of bug reports, assigned to each developer besides automating the triaging problem. These two factors significantly optimize the overall bug fixing time. In this approach, a group of bug reports and developers are represented as nodes in the bipartite graph. The bug reports undergo some text cleaning techniques namely, stop word removal, lemmatization.

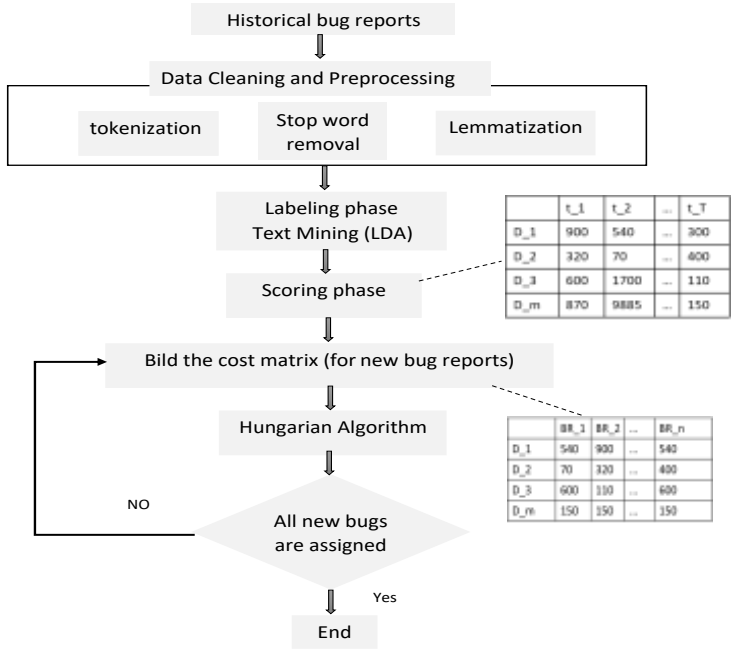

Figure 2: Proposed approach

## A. Fundamental concepts:

### 1) Preprocessing phase:

To apply the labeling technique on the textual data. Some preprocessing steps should be taken using natural language processing (NLP) techniques.

Tokenization: usually, it is the first step in handling the textual data. It is responsible for turning the natural language (unstructured data) into numerical structured information that can be represented as a vector. There are many techniques to apply this step. NLTK (Natural Language Toolkit) is the most common one. It is an open-source library for tokenization. It can handle word and sentence tokenizer, punctuation-based tokenizer and Treebank word tokenizer.

Stop word removal: this is an essential step in pre-processing stage. The point is removing the words such as "a", "an", "the", "how" and "in" that make some noise in the text and this typically affect the processing time. These words are known as not very discriminatory, means that they have no effects on information retrieval and classification. NLTK library is also used to apply this technique.

### 2) Labeling phase:

Topic modeling technique: For the purpose of classifying or labeling the documents in the corpus, Topic modeling is used as an unsupervised natural language processing technique. It is used for the purpose of extracting the hidden topics from a vast number of documents. Mainly, it focuses on the topics (words) that best represent each document in the corpus. It also shows the probability of how each document is related to each topic in percentages. The usual inputs to apply this technique are the number of topics and the corpus documents. Many algorithms are used in this approach such as Latent Dirichlet allocation (LDA), Latent Semantic Analysis and Probabilistic Latent Semantic Analysis.

Figure 3 is a graphical presentation for the LDA. α is the document-topic density, β is topic- word density and M is number of documents in the whole corpus. While N represents the number of words in each document, θ is topic distribution for document M, z represents the topic assigned to each word w and finally, w is the word in α topic.

α and β are hypered attributes that depends on the corpus documents. Optimal model requires correct values for these two parameters. Large values for α refers to more topics in each document and vise-vera. However, large values for β represent that each topic has more words and vise-versa.

### 3) Assignment phase

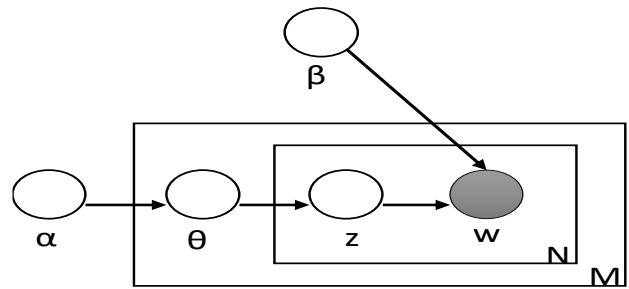

Figure 3: Graphical representation for LDA

Bipartite graph: it is a sort of graph for representing vertices in a form of two independent sets S_1 and S_2. Each node in S_1can be connected to a node in S_2 this link can represent the cost or type of connection between the two nodes. In complete bipartite graph, each vertex in the S_1 is linked to each one in S_2 and vice versa. Figure 5 represents a complete bipartite graph where S_1={A,B,C,D} and S_2={1,2,3}. This graph matching theory mainly used in assignment problems to find the optimal correspondence to execute the goal of the objective function whether it is minimizing or maximizing the edge dependency. The mechanism of bipartite graph depends on finding the subset of edges, such that no vertex in a set share more than one edge. In fact, it is a powerful technique in modeling machine leaning and data mining problems. For example, recommendation systems problems are best described by bipartite graph where S1 represents the items and S2 represents the users. In addition, citation network analysis problems are modeled as bipartite graph, where the publications are the vertices on both sides. Machine learning problems such as question answer mapping and text classification are best described by bipartite graph matching [26].

Table 1: Original Kuhn-Munkres algorithm

| **Original Kuhn-Munkres algorithm** |
|---|
| Input: a bipartite graph $G = (V, E)$ and weight $w(x, y)$. |
| Output: optimal perfect matching $M$. |
| Algorithm steps: |
| Step 1: Generate initial labeling $l$ and matching $M$ in $G_l$. |
| Step 2: If $M$ perfect, stop. Otherwise pick free vertex $u \in X$. Set $S = u, T = \emptyset$. |
| Step 3: If $J_l(S) = T$, update labels (forcing $J_l(S) \neq T$) $$\alpha_l = \min_{s \in S, y \notin T}\{l(x) + l(y) - w(x, y)\}$$ $$\hat{l}(v) = \begin{cases} l(v) - \alpha_l, & v \in S \\ l(v) + \alpha_l, & v \in T \\ l(v), & \text{otherwise} \end{cases}$$ |
| Step 4: If $J_l(S) \neq T$, chose $y \in J_l(S) - T$: <br> • If $y$ free, $u - y$ is augmenting path. Augment $M$ and go to 2. <br> • If $y$ matched, say to $z$, extend alternating tree: $S = S \cup z$, $T = T \cup y$. Go to 3. |

Kuhn-Munkres algorithm: It also called the Hungarian algorithm. It is an optimization algorithm that used to solve the assignment problems. It is used in bipartite graphs to find the maximum weight matching in polynomial time. It was originally developed by Harold Kuhn in 1955 and then James Munkres reviewed it in 1957. Its time complexity is o(n^4). However, Edmonds and Karp proved that it could be modified to have complexity of o(n^3). Using the adjacency matrix (a squared matrix which represents the bipartite graph), the Hungarian algorithm works on providing the optimal assignment which represents the least cost. [43].

### B. Experiment setup:

The proposed solution includes four main stages which are: textual preprocessing for the description field in each bug report, labeling each document (bug report) by topic modeling technique, scoring system to set a score for each developer and the assignment phase to allocate a developer for each bug report.

*a) Database study:* Most of issue tracking systems consider recording the time spent in fixing a bug is useless. So, they ignore the factor of effort, how long a developer took to fix the assigned bug report. However, JBoss project in Jira bug repository is an exception. Jira becomes much more popular than Bugzilla because it provides the time-spent factor in most of bug reports. Jira also provides free license for non-profit organizations [20]. However, among many bug reports recorded in JBoss project, relatively a finite number of them has time-spent factor. They will be our main focus. In this paper, we export a number of JBoss project bug reports from Jira bug repository through the RedHat interface [44]. Furthermore, we specify our search to include the issues with "BUG" type, "DONE" resolution and "CLOSED" status to ensure that all bugs are completely fixed. We specified the bug report factors to be only the bug ID, description, assignee and time spent. In total, 37966 bug report met out search specifications. However, among them, only 410 bug report has a value in the time-spent field and finally become the dataset for our work. With # developer.

In addition, because this small number, we include a huge number of bug reports from Bugzilla and we estimate the fixing time as the difference between the creation and closing dates. A very similar calculation for the time is used by Shihab [45]. We used 26,317 bug report with 516 developers.

*b) Applying the algorithm on Bugzilla dataset*

Similar to what Xuan did in his work [46], we removed the records which solved by the developer who fixed less than 10 bug reports. Also, we ignored the bug reports with empty description. Then, we put the records on chronological order in order to split the data into training and testing. The submitting date of the testing dataset should be after the date of the training dataset (historical bugs).

For the purpose of preprocessing, we applied tokenization and lemmatization on the field of description. In this stage each bug report (represented by the description field) is ready for the classification. We applied the topic modeling algorithm to classify the bug reports according to the topic. However, to apply topic modeling, we have to decide the number of topics.

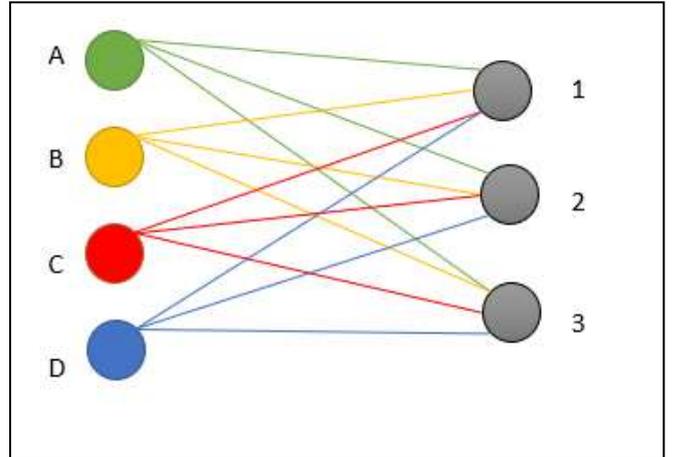

Figure 4: Bipartite graph example

To specify the optimal number of topics to feed the topic modeling algorithm, we developed a part of code that calculate the coherence score for number of topics from 1 to k. Coherence score calculates to what extent words in the same topic make scene together through providing the probabilistic coherence of each number. The higher the score for each specific value of k, the more sense words work together in each topic and the more meaningful the topic will be.

Thus, we used the coherence score result to apply the topic modeling. The result of topic modeling is group of percentage representing how each bug is related to each topic. As an assumption, we label each bug by the topic which has the maximum relationship (maximum percentage).

Using the training dataset, we set a score for each developer by his average fixing time according to the topic modeling results. Thus, we used Equation 1.

$$S(D)_T = \frac{\text{Sum of avg time spent in topic } T \text{ reports solved by developer } D}{\text{total number of topic } T \text{ reports solved by developer } D} \quad (1)$$

*c) Building the cost matrix*

Cost matrix is a requirement for the Hungarian algorithm. It should not include any negative values. It describes the edges and weights (average time spend) between the two sets: (bug reports) and (developers). So, it shows the average time that a developer can take to fix each bug report in the testing dataset. This average time (the result of the scoring system) is decided by the topic of the bug report. In other words, each cell represents the time that developer D takes to fix the corresponding BR according to the topic by which the BR is labeled. A very important condition in the cost matrix is, it should be square matrix, which means the number of columns (testing bug reports) must be equal to the number of rows (existing developers). To overcome this limitation, we used what is called an iterative Hungarian algorithm.

*d) Applying the iterative Hungarian algorithm*

In our experiment, the number of developers is not equal to the number of bug reports. Thus, the cost matrix is can never be square. So, we cannot make the assignment in one step. Our approach proposes to split the testing dataset into sections, each one should be equal to the number of developers. Figure 5 show how we execute this operation using 5 iterations. In the 5th iteration only a group of developers will be considered according to the number of the last group of bug reports.

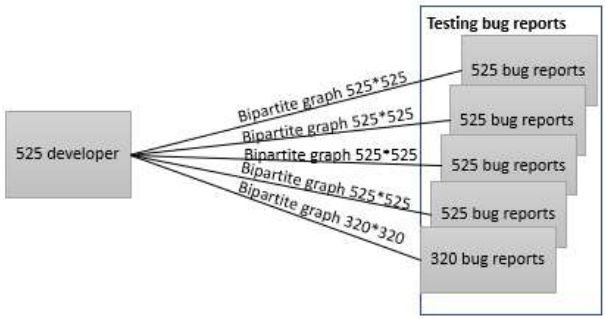

Figure 5: How the iterative Munkres works

## IV. RESULTS AND DISCUSSION

To evaluate our approach, we used the time series split approach. This technique is a special case from the 10-fold cross validation, it is used only in the problems which consider the time factor. Table 2 clearly describe how it split the dataset into training and testing in each iteration. It uses only a part of the dataset and split it into training and testing. The testing dataset remains always the latter 10% of the used datasets while the training part (the first part) increases to be bigger in every iteration.

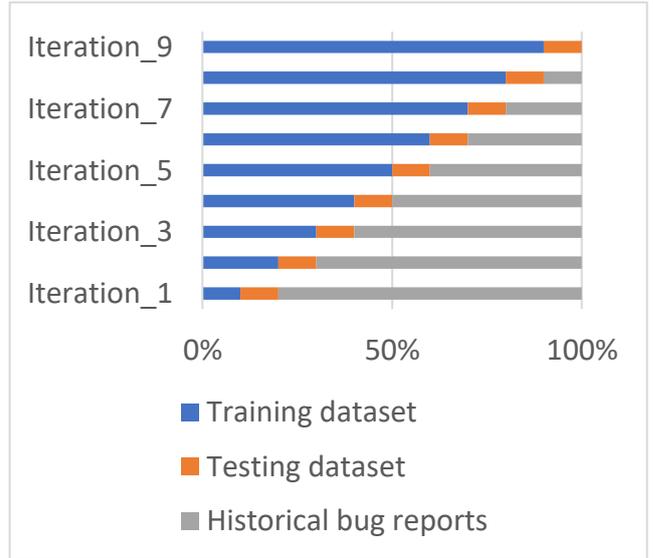

Figure 6: Time series split approach

In our experiment, we specified the range of T (the number of topics to apply the topic modeling) between 1 and 15. By using the time series split approach, the result of applying the coherence method to find the best number of topics will vary in each iteration, which is clear in table 2.

Table 2: Bugzilla Results

| | # Bug reports | # Testing dataset (10%) = 2532 bug report |
|---|---|---|
| Iteration_1 (20% of the dataset) | | # Training dataset (10%) = 2532 bug report |
| | # Developer | 83 developers |
| | Real load | Min =0, Max= 310 |
| | # Topics | 13 |
| | Time reduction | 47.08%. |
| | Resulted developer load | Avg 26-27 bug report for each developer |
| Iteration_2 (30% of the dataset) | # Bug reports | # Testing dataset (10%) = 2532 bug report |
| | | # Training dataset (20%) = 5064 bug report |
| | # Developer | 114 developers |
| | Real load | Min =0, Max= 507 |
| | # Topics | 13 |
| | Time reduction | 15.09% |
| | Resulted developer load | Avg 20-21 bug report for each developer |
| Iteration_3 (40% of the dataset) | # Bug reports | # Testing dataset (10%) = 2532 bug report |
| | | # Training dataset (30%) = 7596 bug report |
| | # Developer | 150 developers |
| | Real load | Min =0, Max= 437 |
| | # Topics | 10 |
| | Time reduction | 13.28% |
| | Resulted developer load | Avg 16-17 bug report for each developer |
| Iteration_4 (50% of the dataset) | # Bug reports | # Testing dataset (10%) = 2532 bug report |
| | | # Training dataset (40%) = 10128 bug report |
| | # Developer | 166 developers |
| | Real load | Min =0, Max=468 |

|  | # Topics | 13 |
|---|---|---|
|  | Time reduction | 14.66% |
|  | Resulted developer load | Avg 14-15 bug report for each developer |
| Iteration_5 (60% of the dataset) | # Bug reports | # Testing dataset (10%) = 2532 bug report |
|  |  | # Training dataset (50%) =12660 bug report |
|  | # Developer | 188 developers |
|  | Real load | Min =0, Max=304 |
|  | # Topics | 14 |
|  | Time reduction | 17.9% |
|  | Resulted developer load | Avg 13-14 bug report for each developer |
| Iteration_6 (70% of the dataset) | # Bug reports | # Testing dataset (10%) = 2532 bug report |
|  |  | # Training dataset (60%) = 15192bug report |
|  | # Developer | 206 developers |
|  | Real load | Min =0, Max=425 |
|  | # Topics | 10 |
|  | Time reduction | 10.65% |
|  | Resulted developer load | Avg 12-13 bug report for each developer |
| Iteration_7 (80% of the dataset) | # Bug reports | # Testing dataset (10%) = 2532 bug report |
|  |  | # Training dataset (70%) =17724 bug report |
|  | # Developer | 226 developers |
|  | Real load | Min =0, Max=418 |
|  | # Topics | 14 |
|  | Time reduction | 16.4% |
|  | Resulted developer load | Avg 11-12 bug report for each developer |
| Iteration_8 (90% of the dataset) | # Bug reports | # Testing dataset (10%) = 2532 bug report |
|  |  | # Training dataset (80%) = 20256 bug report |
|  | # Developer | 240 developers |
|  | Real load | Min =0, Max=390 |
|  | # Topics | 14 |
|  | Time reduction | 13.31 |
|  | Resulted developer load | Avg 30-33 bug report for each developer |
| Iteration_9 (100% of the dataset) | # Bug reports | # Testing dataset (10%) = 2532 bug report |
|  |  | # Training dataset (90%) = 22788bug report |
|  | # Developer | 221 developers |
|  | Real load | Min =0, Max=473 |
|  | # Topics | 14 |
|  | Time reduction | 5% |
|  | Resulted developer load | Avg 11-12 bug report for each developer |

### A. Result analysis

Table 2 shows that the iterative Kuhn-Munkres algorithm successfully optimized the total fixing time by more than 17% of the original fixing time. On top of that, the comparison between the developer work load in manual triage and our automatic one clearly proves the effectiveness of our approach.

## V. CONCLUSION

To put it in a nutshell, bug triaging cos is highly affected by the way we assign the bug report to developers. Unlike the manual triage, which consumes time and resources and put heavy workload on a specific group of developers, our framework is designed and implemented with the intension to solve both of these problems. It normalizes the developer's workload and triage the bugs in a way that optimizes the total fixing time of the bug reports.